\documentclass[a4paper,11pt]{article}
\usepackage{pos}

\title{Four-dimensional QCD equation of state at finite chemical potentials}

\author*[a]{Akihiko Monnai}
\author[b]{Gr\'{e}goire Pihan}
\author[c]{Bj\"orn Schenke}
\author[d]{Chun Shen}

\affiliation[a]{Department of General Education, Faculty of Engineering, Osaka Institute of Technology, \\
Osaka 535-8585, Japan}

\affiliation[b]{Physics Department, University of Houston, \\
Houston, Texas 77004, USA}

\affiliation[c]{Physics Department, Brookhaven National Laboratory, \\
Upton, NY 11973, USA}

\affiliation[d]{Department of Physics and Astronomy, Wayne State University, \\
Detroit, Michigan 48201, USA}

\emailAdd{akihiko.monnai@oit.ac.jp}

\abstract{Exploration of the QCD phase diagram is pivotal in particle and nuclear physics. We construct a full four-dimensional equation of state of QCD with net baryon, electric charge, and strangeness by extending the \textsc{neos} model beyond the conventional two-dimensional approximation. Lattice QCD calculations based on the Taylor expansion method and the hadron resonance gas model are considered for the construction. We also develop an efficient numerical method for applying the four-dimensional equation of state to relativistic hydrodynamic simulations, which can be used for the analysis of nuclear collisions at beam energy scan energies and for different nuclear species at the BNL Relativistic Heavy Ion Collider. }

\FullConference{The XVIth Quark Confinement and the Hadron Spectrum Conference (QCHSC24)\\
 19-24 August, 2024\\
 Cairns Convention Centre, Cairns, Queensland, Australia\\}

\begin{document}
\maketitle

\section{Introduction}

Nuclear matter under extreme conditions has been a topic of significant interest. First-principles calculations based on the lattice QCD method indicate that quarks and gluons are deconfined from hadrons to form the quark-gluon plasma (QGP) at high temperatures above 155-160~MeV \cite{Borsanyi:2013bia,Bazavov:2014pvz,Bollweg:2022fqq}. Nuclear collisions at the BNL Relativistic Heavy Ion Collider (RHIC) and the CERN Large Hadron Collider (LHC) can experimentally produce the QGP, providing unique opportunities to explore the dynamical evolution and statistical properties of the QCD system at various energies with different colliding nuclei. The beam energy scan (BES) program at RHIC is devoted to the exploration of the QCD phase diagram \cite{STAR:2010vob}.

A major discovery in high-energy nuclear collisions is that the QGP exhibits near-perfect fluidity \cite{Kolb:2000fha,Schenke:2010rr}. The relativistic hydrodynamic model with hadronic afterburner and realistic initial conditions has been successful in describing the centrality, collision energy, and momentum dependencies of hadronic yields. It has been established as a unique tool for describing the strongly coupled QCD system dynamically near the quark-hadron crossover temperature, where first-principles methods face difficulties. The model is supplemented with the equation of state, in which the thermodynamic properties of the system are encoded. 

One of the goals of QCD phenomenology is to elucidate the QCD equation of state at finite densities  \cite{Huovinen:2009yb,Moreland:2015dvc,Parotto:2018pwx,Monnai:2019hkn,Noronha-Hostler:2019ayj,Auvinen:2020mpc,Monnai:2021kgu,Kahangirwe:2024cny,Plumberg:2024leb,Monnai:2024pvy}. In this study, we develop a lattice-QCD-based model for the equation of state with multiple conserved charges: net baryon (B), electric charge (Q), and strangeness (S) \cite{Werner:2010aa,Monnai:2019hkn,Noronha-Hostler:2019ayj,Aryal:2020ocm,Karthein:2021nxe,Schafer:2021csj,Monnai:2021kgu}. The hadron resonance gas model, which can reproduce lattice QCD results at zero chemical potentials, is employed as a description of the low-temperature regime. Such models can be found in Refs.~\cite{Monnai:2019hkn,Noronha-Hostler:2019ayj} as functions of temperature and chemical potentials, but they have not been fully formulated as functions of energy and conserved charge densities, which appear in the hydrodynamic equations of motion. We introduce novel variables that enable evolution of multiple conserved currents and efficient numerical simulations of relativistic nuclear collisions \cite{Pihan:2023dsb,Pihan:2024lxw}. An alternative approach can be found in Refs.~\cite{Plumberg:2023vkw,Plumberg:2024leb}.

\section{The equation of state}

The hybrid QCD equation of state model, \textsc{neos-4d}, is constructed based on the Taylor expansion method \cite{Gavai:2001fr,Allton:2002zi} of lattice QCD simulations:
\begin{equation}
\frac{P_\mathrm{lat}}{T^4} = \frac{P_0}{T^4} + \sum_{l,m,n} \frac{\chi^{B,Q,S}_{l,m,n}}{l!m!n!} \bigg( \frac{\mu_B}{T} \bigg)^{l}  \bigg( \frac{\mu_Q}{T} \bigg)^{m}  \bigg( \frac{\mu_S}{T} \bigg)^{n}. 
\label{eq:P_lat}
\end{equation}
$P_0$ and $\chi^{B,Q,S}_{l,m,n}$ are the pressure and susceptibilities defined at vanishing chemical potentials. We use all susceptibilities up to the fourth order from lattice QCD simulations \cite{Bazavov:2014pvz,Bazavov:2012jq, Ding:2015fca, Bazavov:2017dus, Sharma} and parametrize $\chi_{6}^B$, $\chi_{51}^{BQ}$, and $\chi_{51}^{QS}$ as defined in Ref.~\cite{Monnai:2019hkn} for thermodynamic consistency.
The pressure is smoothly connected to that of the hadron resonance gas model around the quark-hadron crossover because the expansion method is not convergent at large fugacities \cite{Karsch:2010hm}:
\begin{align}
P &= \frac{1}{2}\bigg(1- \tanh \frac{T-T_c}{\Delta T_c}\bigg) P_{\mathrm{had}} + \frac{1}{2}\bigg(1+ \tanh \frac{T-T_c}{\Delta T_c}\bigg) P_{\mathrm{lat}}. \label{eq:P}
\end{align}
This yields a crossover equation of state. $T_c(\mu_B) = a - d  (b \mu_B^2 + c \mu_B^4)$ is the connection temperature and $\Delta T_c = 0.1T_c (0)$ is the width of the connection region. We use $a = 0.16\ \mathrm{GeV}$, $b = 0.139\ \mathrm{GeV}^{-1}$, $c = 0.053\ \mathrm{GeV}^{-3} $, and $d = 0.4$ motivated by the chemical freezeout \cite{Cleymans:2005xv}. The hadronic pressure is 
\begin{equation}
P_\mathrm{had} = \pm T \sum_i \int \frac{g_i d^3p}{(2\pi)^3} \ln [1 \pm e^{-(E_i-\mu_i)/T}], \label{eq:P_had}
\end{equation}
where the index $i$ denotes hadronic particle species. We consider the hadrons and resonances lighter than 2 GeV that have $u$, $d$, or $s$ as valence quarks from the particle data group \cite{Tanabashi:2018oca}. $g_i$ is the degeneracy and $\mu_i$ is the hadronic chemical potential. The positive sign is for baryons and the negative sign is for mesons. Other thermodynamic variables, such as $e$, $s$, $n_B$, $n_Q$, $n_S$, and $c_s$, can be obtained through the fundamental thermodynamic relations.

\section{Numerical results}

\begin{figure}[tb]
\centering
\includegraphics[width=2.8in]{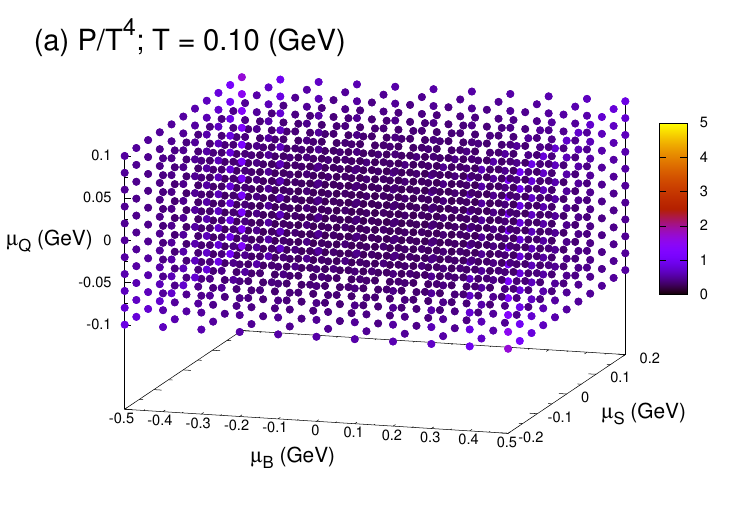}%
\includegraphics[width=2.8in]{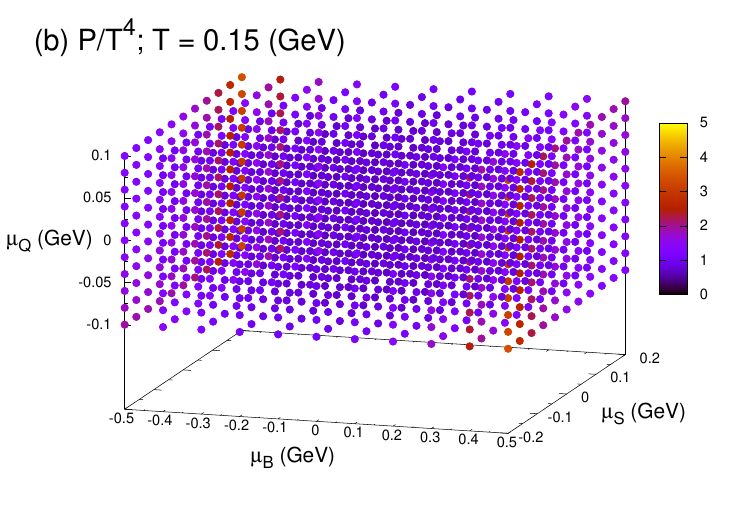}
\includegraphics[width=2.8in]{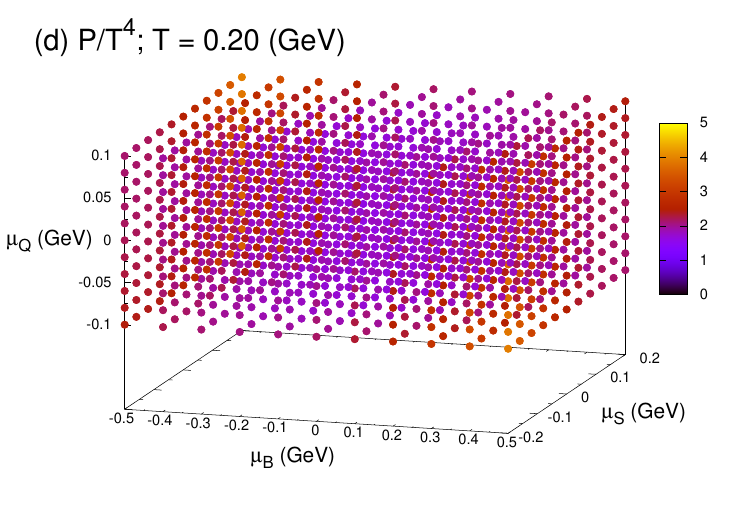}%
\includegraphics[width=2.8in]{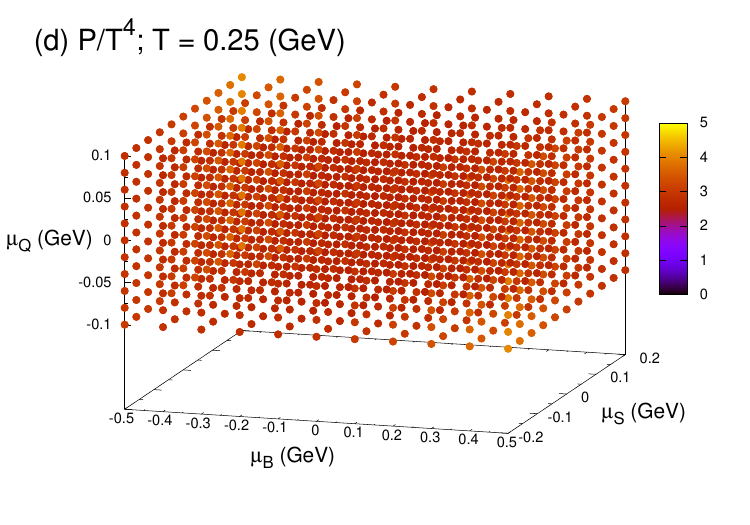}
\includegraphics[width=2.8in]{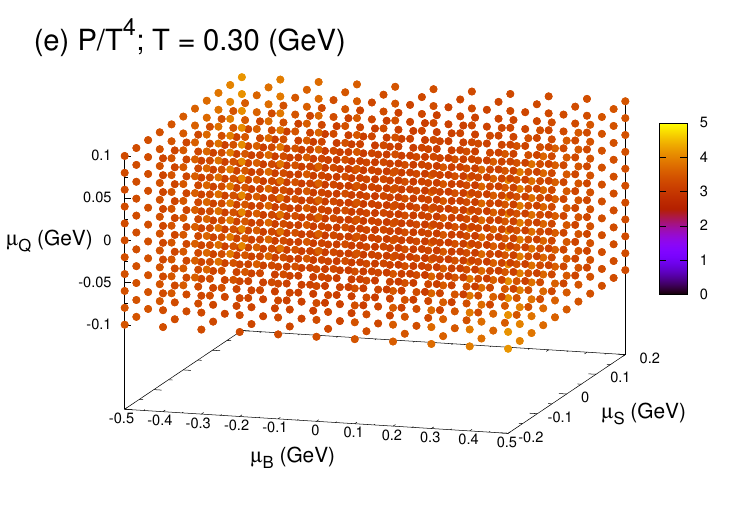}%
\includegraphics[width=2.8in]{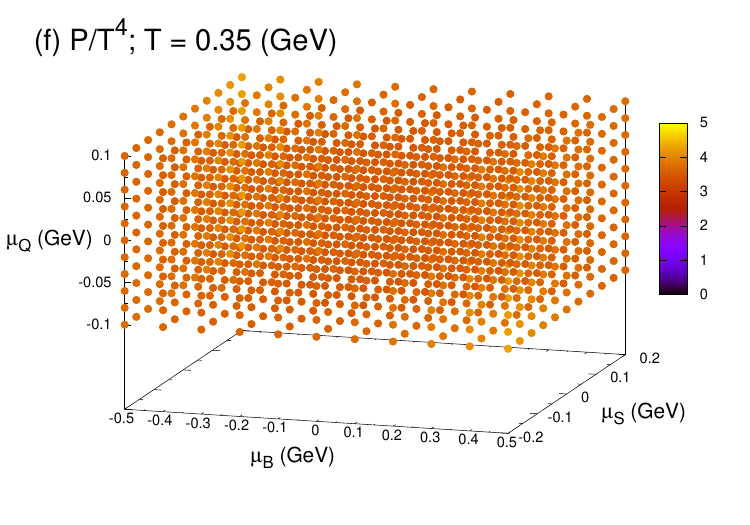}
\caption{Dimensionless pressure $P/T^4$ as function of $\mu_B,\mu_Q$, and $\mu_S$ at (a) $T=0.10$ GeV, (b) $T=0.15$ GeV, (c) $T=0.20$ GeV, (d) $T=0.25$ GeV, (e) $T=0.30$ GeV, and (f) $T=0.35$ GeV.}
\label{fig:pressure}
\end{figure}

\begin{figure}[tb]
\centering
\includegraphics[width=2.8in]{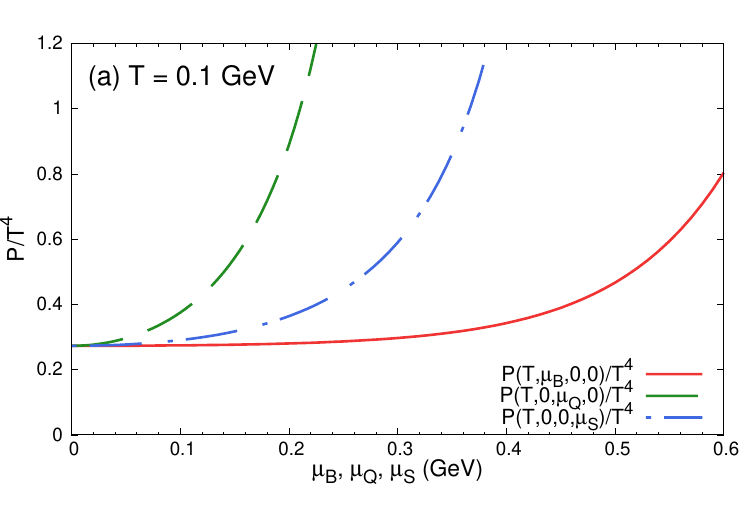}
\includegraphics[width=2.8in]{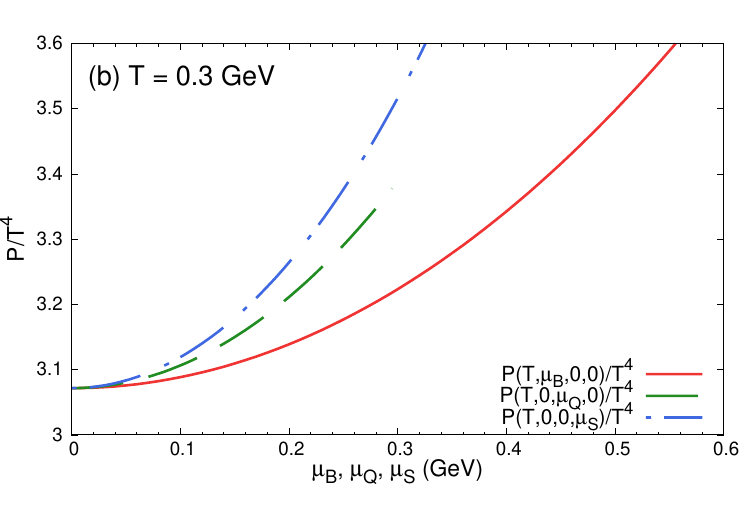}
\caption{Dependence of dimensionless pressure $P/T^4$ on $\mu_B,\mu_Q$, or $\mu_S$ at (a) $T=0.10$ GeV and (b) $T=0.30$ GeV.}
\label{fig:chem_dep}
\end{figure}

We estimate the four-dimensional equation of state in numerical simulations. Dimensionless pressure $P/T^4$ as a function of $\mu_B$, $\mu_Q$, and $\mu_S$ is shown at different $T$ in the vicinity of the QCD crossover in Fig.~\ref{fig:pressure}. One can see that the pressure tends to become large at larger temperatures and chemical potentials. It is intrinsically symmetric under the simultaneous sign conversion of $\mu_B$, $\mu_Q$, and $\mu_S$. On the other hand, the pressure is asymmetric under the sign conversion of a single chemical potential because off-diagonal susceptibilities mix the contributions of different chemical potentials. 
Figure~\ref{fig:chem_dep} shows the dependence of $P/T^4$ on each chemical potential. It is most sensitive to $\mu_Q$ and least to $\mu_B$ in the hadronic phase because the lightest carriers of electric, strangeness, and baryon charges are pions, kaons, and protons, respectively. On the other hand, in the parton gas limit $P/T^4$ is most sensitive to $\mu_S$ and least to $\mu_B$ because the lowest-order diagonal susceptibilities approach $\chi_2^S = 1$, $\chi_2^Q = 2/3$, and $\chi_2^B = 1/3$ (see also Eqs.(\ref{eq:nb_parton})-(\ref{eq:ns_parton}) for references).

The entropy-to-baryon density ratio $s/n_B$ remains unchanged during ideal hydrodynamic evolution in nuclear collisions. Therefore, the constant $s/n_B$ lines can characterize the region in the phase diagram explored at each collision energy \cite{Gunther:2016vcp}. The trajectories are also affected by $n_Q$ and $n_S$. Here, we consider $n_Q/n_B = 0$ and $n_Q/n_B = 1$ as limiting cases corresponding to neutron-rich and proton-rich domains of the initial geometry in nuclear collisions, respectively, and the strangeness neutrality condition $n_S = 0$ for demonstration. It should be noted that these conditions can be modified in the presence of additional fluctuation or dissipation.

\begin{figure}[tb]
\includegraphics[height=1.87in]{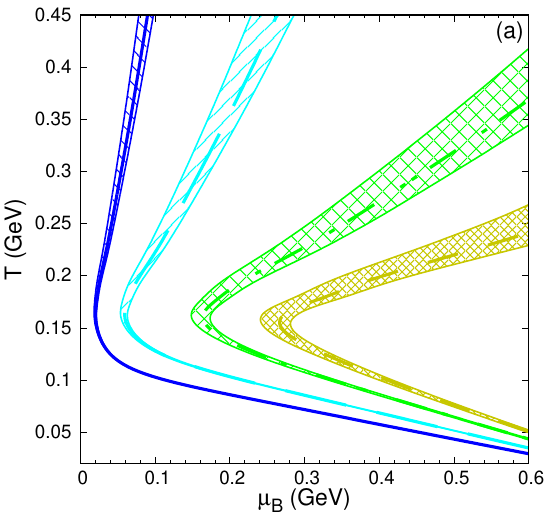}%
\includegraphics[height=1.87in]{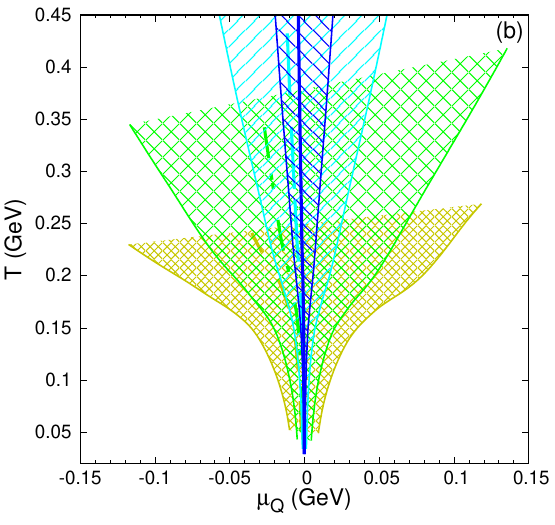}%
\includegraphics[height=1.87in]{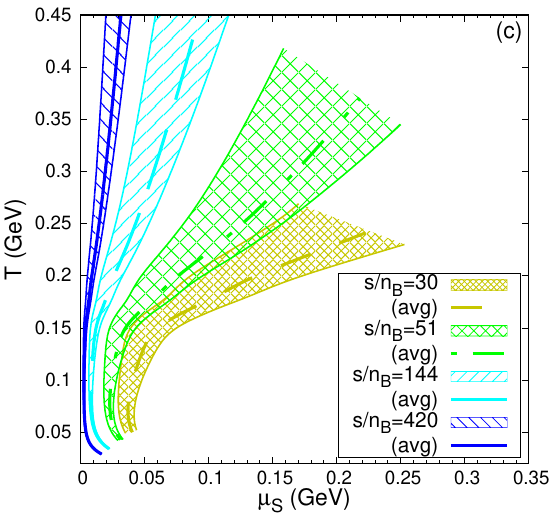}
\caption{The regions probed when $n_Q/n_B$ ranges from 0 to 1 for $s/n_B$ = 420, 144, 51, and 30 on the (a) $T$-$\mu_B$, (b) $T$-$\mu_Q$, and (c) $T$-$\mu_S$ planes. The solid, dashed, dash-dotted, and dotted lines are the average case at $n_Q/n_B$ = 0.4. }
\label{fig:sn}
\end{figure}

Figure~\ref{fig:sn} shows the trajectories of $s/n_B = 420, 144, 51$ and $30$, which represent the collider energies of $\sqrt{s_{\rm NN}}=200, 62.4, 19.6$ and $14.5$ GeV, respectively, in the (a) $T$-$\mu_B$, (b) $T$-$\mu_Q$, and (c) $T$-$\mu_S$ planes, where the other chemical potentials are set to zero. The bands illustrate the range between the $n_Q/n_B = 0$ and $n_Q/n_B = 1$ cases. The thick lines in the middle of the bands denote the average case $n_Q/n_B = 0.4$ that is implied by the $Z/A$ ratio of heavy nuclei such as $^{197}$Au and $^{208}$Pb for reference. They are cut off when $\mu_B$ exceeds 0.6 GeV. The trajectories are bent around the crossover temperature in Fig.~\ref{fig:sn} (a) because baryons are heavier than quarks. As the temperature decreases, the baryon chemical potential needs to increase in the hadronic phase for net baryon number conservation. The average trajectories become slightly negative in  $\mu_Q$ (Fig.~\ref{fig:sn} (b)) because $\mu_n = \mu_B > \mu_p = \mu_B+\mu_Q$ in neutron-rich heavy nuclei. We observe a wide band for $\mu_Q$, implying that a large region in the phase diagram can be explored in relativistic nuclear collisions. $\mu_S$ is finite in Fig.~\ref{fig:sn} (c) owing to the strangeness neutrality condition because the lack of strange quark chemical potential $\mu_s$ leads to $\mu_S = \frac{1}{3}\mu_B - \frac{1}{3}\mu_Q > 0$ when $\mu_B$ is positive and larger than $\mu_Q$. The $s/n_B$ starts to increase below $T=0.05$ GeV, likely because the mass of kaons, which are the primary strangeness carrier there, becomes non-negligible.

Next, we develop an efficient method of application of the equation of state to the hydrodynamic model of nuclear collisions. There are 7 independent equations of motion in inviscid relativistic hydrodynamics with the three conserved charges, $\partial_\mu T^{\mu\nu}=0$ and $\partial_\mu N_{B,Q,S}^{\mu}=0$. They are expressed in terms of 8 independent variables, flow $u^\mu$, energy density $e$, pressure $P$, and conserved charge densities $n_{B,Q,S}$. Thus, one needs the equation of state $P = P(e,n_B,n_Q,n_S)$, which characterizes the thermodynamic properties of the system, to close the set of equations. Since the pressure (\ref{eq:P}) is given as a function of the conjugate variables $T$, $\mu_B$, $\mu_Q$, and $\mu_S$, it needs to be inverted for hydrodynamic simulations. For numerical efficiency, one often prepares pre-calculated tables of the equation of state. However, the tables become too large for standard numerical implementation for a four-dimensional equation of state because $e,n_B,n_Q,n_S$ and $T,\mu_B,\mu_Q,\mu_S$ have different dimensions.

We define new variables $\tilde{T}$, $\tilde{\mu}_B$, $\tilde{\mu}_Q$, and $\tilde{\mu}_S$ as the temperature and chemical potentials of the $N_f = 3$ parton gas with the same energy and conserved charge densities as our system and use them to tabulate the numerical results to overcome this issue. The parton gas equation of state is expressed as
\begin{align}
e &= \frac{19\pi^2}{12} \tilde{T}^4 , \label{eq:e_parton}\\
n_B &= \frac{1}{3} \tilde{\mu}_B \tilde{T}^2 - \frac{1}{3} \tilde{\mu}_S \tilde{T}^2, \label{eq:nb_parton} \\
n_Q &= \frac{2}{3} \tilde{\mu}_Q \tilde{T}^2 + \frac{1}{3} \tilde{\mu}_S \tilde{T}^2, \\
n_S &= - \frac{1}{3} \tilde{\mu}_B \tilde{T}^2 + \frac{1}{3} \tilde{\mu}_Q \tilde{T}^2 + \tilde{\mu}_S \tilde{T}^2 ,\label{eq:ns_parton}
\end{align}
whose solutions can be obtained analytically as
\begin{align}
\tilde{T}(e,n_B,n_Q,n_S) &= \bigg(\frac{12}{19\pi^2} e \bigg)^{1/4} ,\label{eq:tilde_T} \\ 
\tilde{\mu}_B(e,n_B,n_Q,n_S) &= \frac{5 n_B-n_Q+2n_S}{\tilde{T}^2}, \label{eq:tilde_muB}\\
\tilde{\mu}_Q(e,n_B,n_Q,n_S) &= \frac{- n_B+2n_Q-n_S}{\tilde{T}^2}, \label{eq:pg-q}\\
\tilde{\mu}_S(e,n_B,n_Q,n_S) &= \frac{2 n_B - n_Q + 2n_S}{\tilde{T}^2}.\label{eq:tilde_muS}
\end{align}
The variables from the conservation laws can be converted into the temperature and chemical potentials of the parton gas using the relations (\ref{eq:tilde_T})-(\ref{eq:tilde_muS}), and they can be used to access the numerical results tabulated in the form of $P = P(\tilde{T},\tilde{\mu}_B,\tilde{\mu}_Q,\tilde{\mu}_S)$. Similarly, the temperature and chemical potentials can be estimated as functions of the parton gas-based variables for particle production \cite{Cooper:1974mv}. A successful application of the method can be found in Ref.~\cite{Pihan:2024lxw}.

\begin{figure}[tb]
\centering
\includegraphics[width=2.8in]{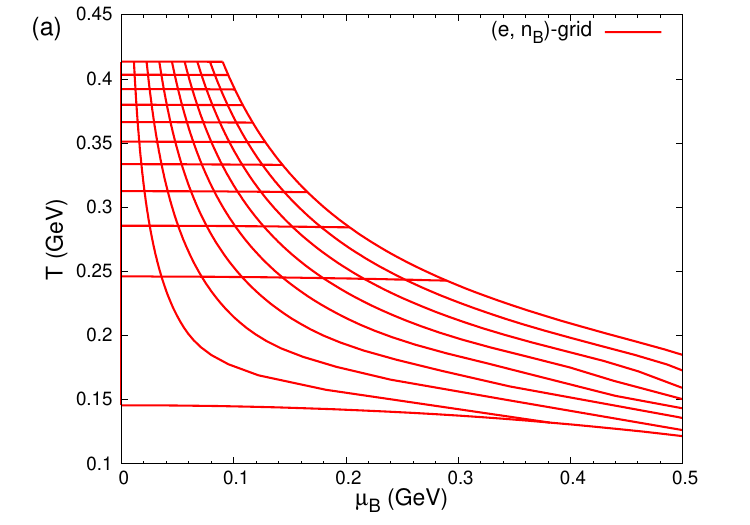}%
\includegraphics[width=2.8in]{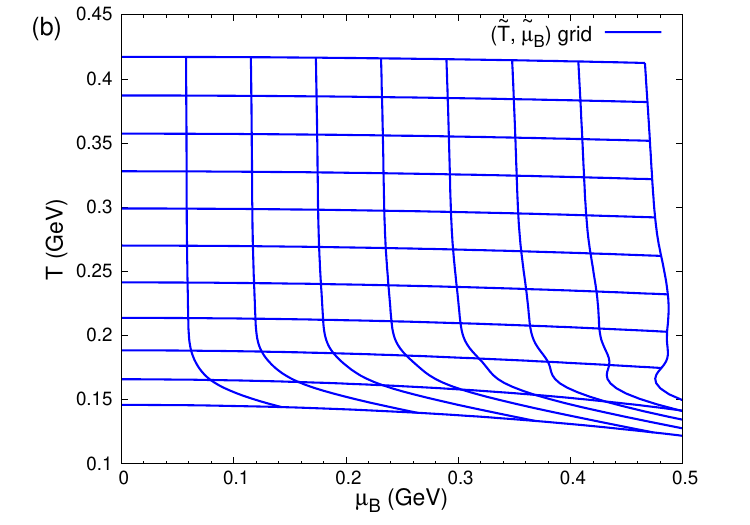}
\caption{Comparison of grids with equal spacing in (a) $e$ and $n_B$ and (b) $\tilde{T}$ and $\tilde{\mu_B}$ on the $T$-$\mu_B$ plane.}
\label{fig:grid}
\end{figure}

The grids with equal spacing in (a) $e$ and $n_B$ and (b) $\tilde{T}$ and $\tilde{\mu_B}$ near the crossover region in the $T$-$\mu_B$ plane at $\mu_Q = \mu_S = 0$ are shown in Fig.~\ref{fig:grid} to illustrate the situation. One can see that the grid is deformed in the conventional $e$-$n_B$ grid, leading to difficulties in covering the whole region especially in the higher dimensional cases. On the other hand, the grid is relatively uniform in the $\tilde{T}$-$\tilde{\mu_B}$ space, allowing more efficient tabulation of the equation of state for numerical hydrodynamic simulations.

\section{Conclusions and outlook}

We have constructed a full four-dimensional model of the QCD equation of state at finite baryon, electric charge, and strangeness densities, \textsc{neos-4d}, for hydrodynamic simulations of relativistic nuclear collisions. The pressures of the lattice QCD simulations in the Taylor expansion method and the hadron resonance gas model are matched near $T_c$ to obtain a crossover equation of state. The other state variables can be obtained using the fundamental thermodynamic relations. The model is implemented numerically and verified to be thermodynamically consistent.

The ratio of conserved charges $n_Q/n_B$ can vary to from 0 to 1, corresponding to neutron-rich and proton-rich regions of the initial geometry. The former may occur in the peripheral collisions of nuclei with neutron skins and the latter in proton-proton collisions. The constant $s/n_B$ trajectories suggest that a wider region in the phase diagram can be explored in the beam energy scan programs, especially in the $\mu_Q$ direction, when the variation is taken into account.

The hydrodynamic equations of motion involve the densities $e$, $n_B$, $n_Q$ and $n_S$. We have introduced the temperature and chemical potentials of a parton gas at the given energy and conserved charge densities, $\tilde{T}$, $\tilde{\mu}_B$, $\tilde{\mu}_Q$, and $\tilde{\mu}_S$, and used them to construct the tables to enable practical and efficient implementation of the equation of state.
Our results serve as a pivotal component in numerical hydrodynamic simulations of nuclear collisions across various collision energies and different nuclear species. The tabulated version of the \textsc{neos-4d} results are publicly available \cite{neos4d}. 

\begin{acknowledgments}
The authors thank Frithjof Karsch, Swagato Mukherjee, and Sayantan Sharma for providing the lattice QCD data. This work is supported by JSPS KAKENHI Grant Numbers JP19K14722 and JP24K07030 (A.M.), by the U.S. Department of Energy, Office of Science, Office of Nuclear Physics, under DOE Contract No.~DE-SC0012704 and within the framework of the Saturated Glue (SURGE) Topical Theory Collaboration (B.P.S.) and Award No.~DE-SC0021969 (C.S. \& G.P.). C.S. acknowledges a DOE Office of Science Early Career Award. 
\end{acknowledgments}

\bibliographystyle{JHEP}
\bibliography{neos}

\end{document}